# Machine Learning-Enabled Mechanical Analysis and Optimization of Bioinspired Functionally Graded Materials

Zhangke Yang[1] and Zhaoxu Meng[1*]

[1] Department of Mechanical Engineering, Clemson University, Clemson, SC, 29634, USA

Corresponding author:

* zmeng@clemson.edu

# Abstract

Tendon-bone enthesis connects tendon and bone, two mechanically dissimilar materials, while effectively minimizing stress concentrations, a capability rarely achieved in engineering materials. Its hierarchical organization and graded variations in composition or mineralization are widely recognized as key contributors to its exceptional performance. Here, we investigate the mechanics of enthesis, focusing on the insertion of interface collagen fibers into bone where hierarchical collagen fibril structures and graded mineralization are present, and translate these insights into bioinspired engineering material design using a convolutional neural network-based field predictor (CNNFP). We first construct a three-dimensional finite element model (FEM) of the interface fiber-bone enthesis, in which local material properties depend on mineralization level, mean fibril orientation, and angular dispersion, informed by a multiscale continuum theory. We introduce a scalar risk factor that integrates local stress states and constituent fibril organizations to quantify local vulnerability. Simulation results demonstrate that graded and spatially heterogeneous configurations markedly reduce stress concentrations, supporting prevailing biomechanical hypotheses. We then train the CNNFP as an accurate surrogate for FEM and embed it within a kernel-based gradient optimization framework to efficiently identify optimal field configurations. The optimized designs are validated against FEM ground truth, establishing a generalizable AI-enabled pathway for the optimization of bioinspired functionally graded materials.

# Introduction

Over millions of years of evolution and natural selection, biological systems have developed materials and tissues with functionally graded or heterogeneous compositions and structures, enabling them to achieve superior properties, such as load-bearing capacity, contact damage resistance, interfacial strengthening, and enhanced toughness, with very limited choices of elements[1]. Biological systems commonly exhibit spatially variations in the content, arrangement, size, and orientation of their fundamental constituents, such as biominerals and collagenous fibrils, tailored to meet functional demands with maximized efficiency [2,3]. In particular, the tendon-bone enthesis is a crucial functional region that connects tendon to bone and transfers mechanical loads between them [4,5]. Bone and tendon are dissimilar biological tissues with significantly different mechanical properties. Specifically, bone is stiff and predominantly isotropic, while tendon is compliant and largely anisotropic [6-8]. Typically, the interface between two adhered materials with vastly different mechanical properties is expected to experience intense stress concentration, therefore prone to failure [9]. Tendon-bone enthesis has effectively addressed this challenge, functioning seamlessly for decades in most people and animals without any issues. Therefore, it has garnered significant interest from researchers seeking to understand the underlying mechanisms, which could offer valuable guidance for engineering design.

Experimental research reveals that tendon-bone enthesis features a highly hierarchical fibrillar structure, primarily composed of collagen fibrils formed by staggered tropocollagen molecules [10-12]. These collagen fibrils are further organized into collagen fibers at the microscale [13,14]. At the sub-millimeter scale, the collagen fibers with different degrees of mineralization (typically dependent on the location) are arranged into what are known as tendon fibers (close to the tendon end) and interface fibers [14-16]. Each tendon fiber originates from the tendon side, branching into multiple interface fibers as it approaches and ultimately attaches to the bone [17,18]. The branching of tendon fibers into multiple interface fibers is one of the two most prominent structural features of tendon-bone enthesis. Our previous work has provided insights into the critical role of the branching architecture in mechanics and failure of the enthesis [19].

Another key structural feature is the gradient in angular dispersion of collagen fibrils relative to the mean fibril orientation. Specifically, the angular dispersion is relatively small near the tendon side, indicating closely aligned collagen fibrils. The angular dispersion increases as it approaches the bone side, indicating more poorly aligned collagen fibrils [16,17,20]. Similar to the gradient in angular dispersion, there is also a gradient in the degree of mineralization within the tendon-bone enthesis, being relatively low near the tendon side and higher near the bone side, which is recognized to be the key compositional feature contributing to the exceptional mechanical properties of enthesis [21-24].

Motivated by the distinct structural and compositional features of tendon-bone enthesis, substantial efforts have been devoted to elucidating its structure-composition-function relationships. Early modeling work by Thomopoulos et al. developed an idealized two-dimensional mechanical model of the enthesis to examine how collagen fiber orientation influences stress concentrations, treating fibers as nearly uniaxial elements governed by an orientation-dependent probability distribution [16]. Building on this concept, Genin et al. introduced a mathematical model incorporating continuous gradients in the orientation distribution of collagen fibers and organic-inorganic composition [6], demonstrating that spatial variations in axial and shear moduli can be directly linked to micrometer-scale tissue features. More recent studies have combined experiments, simulations, advanced imaging, and *in vivo* manipulation of mineralization and architecture to uncover mechanisms underlying the mechanical function of enthesis. Golman et al. revealed trade-offs between strength and toughness within the enthesis and showed that optimal performance emerges from cross-scale architectural adaptations [5]. Rossetti et al. uncovered the microscopic mechanisms governing the physiological function of the enthesis by integrating micromechanical, structural, compositional, and proteomic characterization methods [14]. At an even finer scale, Deymier et al. employed micrometer-resolution mechanical testing and imaging to identify a compliant zone near the mineralization gradient, which is believed to enhance toughness through elevated energy dissipation [25].

Despite extensive investigations into the mechanics of tendon-bone enthesis, comparatively little has been devoted to mesoscale tendon and interface fibers, particularly with respect to their underlying collagen fibril structure, orientation, and compositional heterogeneity. Our previous work has demonstrated the mechanical significance of branching fiber architectures and region-specific mineralization within the enthesis [19]. However, how graded fibril organization and mineralization evolve at the level of individual interface collagen fibers remains largely unresolved. Moreover, it remains unclear how such fiber-level insights can be translated into rational design principles for heterogeneous or gradient-engineered materials that emulate the natural transition from compliant soft tissue to stiff bone.

Individual tendon and interface collagen fibers are themselves hierarchical composites, consisting of collagen fibrils with varying orientations and degrees of mineralization [26,27]. Consequently, the effective mechanical behaviors of individual fibers are governed by the intrinsic properties of their constituent collagen fibrils, their orientation distribution, and the mineralization levels across different length scales [22,23,27,28]. Establishing these structure–property relationships, however, is inherently challenging. Previous studies have shown that the tissue-level effective stiffness tensor can be derived by integrating over the probability density function of fiber orientations [6]. In addition, the mechanical response of an individual collagen fibril depends on the relative fractions of its constituents, including collagen molecules, non-collagenous proteins, and embedded hydroxyapatite crystals [27,29].

To address these multiscale challenges, we employ a hierarchical modeling framework. Specifically, we first establish the effective mechanical behavior of collagen fibrils using a matrix–inclusion approach, treating the fibril as a sequence of inclusion problems across length scales[27]. Building upon this foundation, we construct a three-dimensional finite element model (FEM) of the interface fiber–bone system, in which the local material properties of the interface fiber and adjacent bone explicitly depend on the spatially varying mineralization level, mean fibril orientation, and fibril angular dispersion. This framework enables direct linkage between fibril-scale composition and orientation and the mechanical response of individual interface fibers near the bone end, thereby facilitating a fiber-level interrogation of graded mineralization and fibril organization.

Beyond predicting elastic stress fields using FEM, we introduce a failure metric that explicitly incorporates both local stress states and fibril orientation, yielding a scalar risk factor field that assesses spatial variations in failure susceptibility. To systematically explore these high-dimensional structure–property relationships and translate biological insights into design principles, we further develop a machine-learning (ML) framework that integrates a convolutional neural network [30] for field prediction, i.e., a convolutional neural network-based field predictor (CNNFP), a kernel-based field generator, and an efficient optimization algorithm. This integrated approach enables efficient evaluation of the coupled effects of mineralization and fibril orientation on fiber mechanics and facilitates the identification of optimal field distributions that minimize failure risk. Collectively, this framework provides a scalable, transferable strategy for analyzing natural tendon–bone interfaces and guiding the rational design of heterogeneous, functionally graded materials.

## Results

**Mechanical analysis via computational modeling of the interface fiber-bone system**

FEM of a representative volume element of the fiber-bone interface system is developed to capture the essential structural features of the enthesis (Fig. 1A). To balance fidelity and computational efficiency, only a localized region encompassing the interface fiber insertion into bone is considered. This region is selected because it is mechanically critical and prone to stress concentration due to the sharp contrast in material properties between soft and hard tissues.

As illustrated in Fig. 1B, the model consists of two cylindrical domains: a smaller cylinder representing the interface fiber segment and a larger cylinder representing the adjacent bone. The interface fiber is modeled as a cylinder with a height and diameter of 100 $\mu m$, consistent with experimental observations of porcine Achilles tendon-bone enthesis [14]. The bone domain is represented by a larger cylinder with a height of 50 $\mu m$ and a diameter of 200 $\mu m$, with its bottom surface constrained to approximate the mechanical confinement imposed by the surrounding bulk bone. To reflect the smooth transition observed experimentally at tendon–bone enthesis [5,14], while maintaining generality, a fillet with a radius of 30 $\mu m$ is introduced at the fiber-bone interface. Notably, similar smooth transitions are also observed at ligament–bone entheses, cartilage–bone junctions, and other fibrocartilaginous soft tissue–bone interfaces [31-33], underscoring the broader relevance of this geometric representation.

The top surface of the interface fiber is subjected to normal, shear, or combined loading conditions, depending on the specific case studied. The mechanical behavior of the interface fiber and bone is governed by the hierarchical organization of their constituent collagen fibrils, specifically the local mineralization level and fibril orientation distribution [6,26,27]. To capture these effects, we adopt a previously developed multiscale continuum theory to formulate the constitutive behavior of the interface fiber with varying fibril structure and mineralization, as detailed in the **Methods** Section.

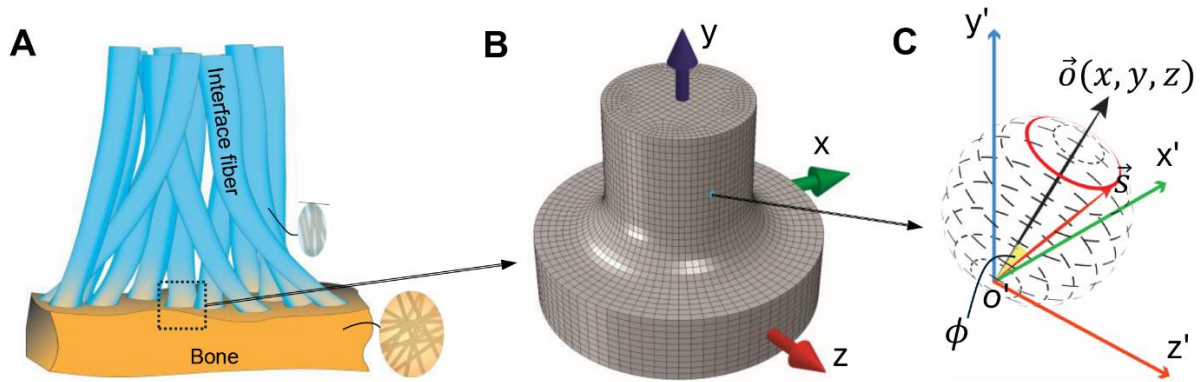

**Figure 1 | Continuum finite element model of a representative interface fiber at the tendon-bone enthesis. A.** Schematic illustration of interface fibers inserting into bone, with insets illustrating the underlying collagen fibrillar structure. **B.** 3D FEM represents the geometric transition from the interface fiber to the bone. **C.** Definition of the local coordinate system, where $z'$ denotes the radial direction and $y'$ is aligned with the global $y$-axis.

To evaluate the mechanical response under physiologically representative conditions, we apply a uniform tensile stress of 50 MPa along the fiber axis (positive y-direction), while constraining the bottom surface of the bone. In addition to analyzing the 3D stress fields, we introduce an element-based risk factor that integrates the local 3D stress state with the orientational distribution of collagen fibrils at each spatial point, providing a more effective scalar measure of failure susceptibility. The specific definition of the risk factor is detailed in the **Methods** Section. Implicit static simulations are performed to compute stress fields and corresponding risk factor distributions. Owing to the axisymmetric geometry, boundary conditions, and fibril orientation and mineralization scale fields (details in the **Methods** Section), the resulting solutions are also axisymmetric. Accordingly, both input and output fields are presented and visualized on axisymmetric planes.

**Effect of gradient mineralization scale and fibril angular dispersion on enthesis mechanics**

We examine the influence of gradient mineralization scale and fibril angular dispersion, two defining features of natural tendon–bone enthesis. Experimental studies indicate that both mineralization level and angular dispersion are relatively low near the tendon side and progressively increase toward the bone side [14,17]. Although the precise spatial evolution of these features varies among tissues and anatomical locations, our baseline model assumes linear increases in both mineralization and angular dispersion along the tendon-to-bone direction (Fig. 2B, D). For comparison, we also consider an extreme, non-physiological case in which the tendon is directly connected to the bone without any transitional gradients, represented by the discontinuous fields shown in Fig. 2C, E. To isolate the mechanical effects of mineralization and angular dispersion, we prescribe a constant mean fibril orientation aligned with the fiber axis (y-direction) across all cases (Fig. 2F–H). This assumption closely experimentally observed fibril alignment in tendon–bone entheses [4,17,29], allowing us to focus specifically on the role of gradient fields in governing fiber-level mechanics.

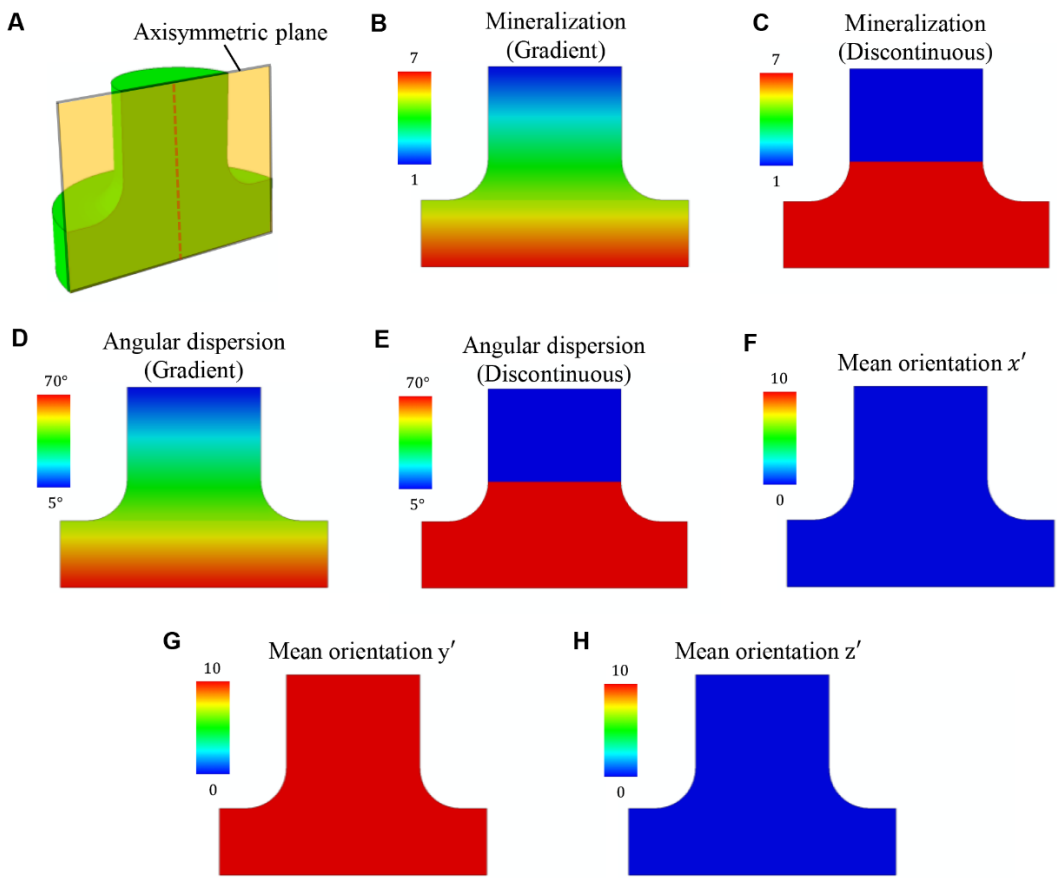

**Figure 2 | Input fields for gradient vs. discontinuous mineralization scale and angular dispersion. A.** Schematic of the axisymmetric plane on which input and output fields are visualized. **B.** Gradient mineralization scale field. **C.** Discontinuous mineralization scale field. **D.** Gradient fibril angular dispersion field. **E.** Discontinuous fibril angular dispersion field. **F-H.** Component fields of the mean fibril orientation in the $x'$, $y'$, and $z'$ directions, respectively.

Fig. 3 presents axisymmetric contours of stress components and the element-based risk factor under discontinuous versus gradient mineralization scale and fibril angular dispersion fields. Specifically, Fig. 3A-H show 3D stress components, the von Mises stress, and the risk factor, respectively, for discontinuous fields, while Fig. 3I-P display the corresponding results of the gradient case. All stress components are reported in the local coordinate system.

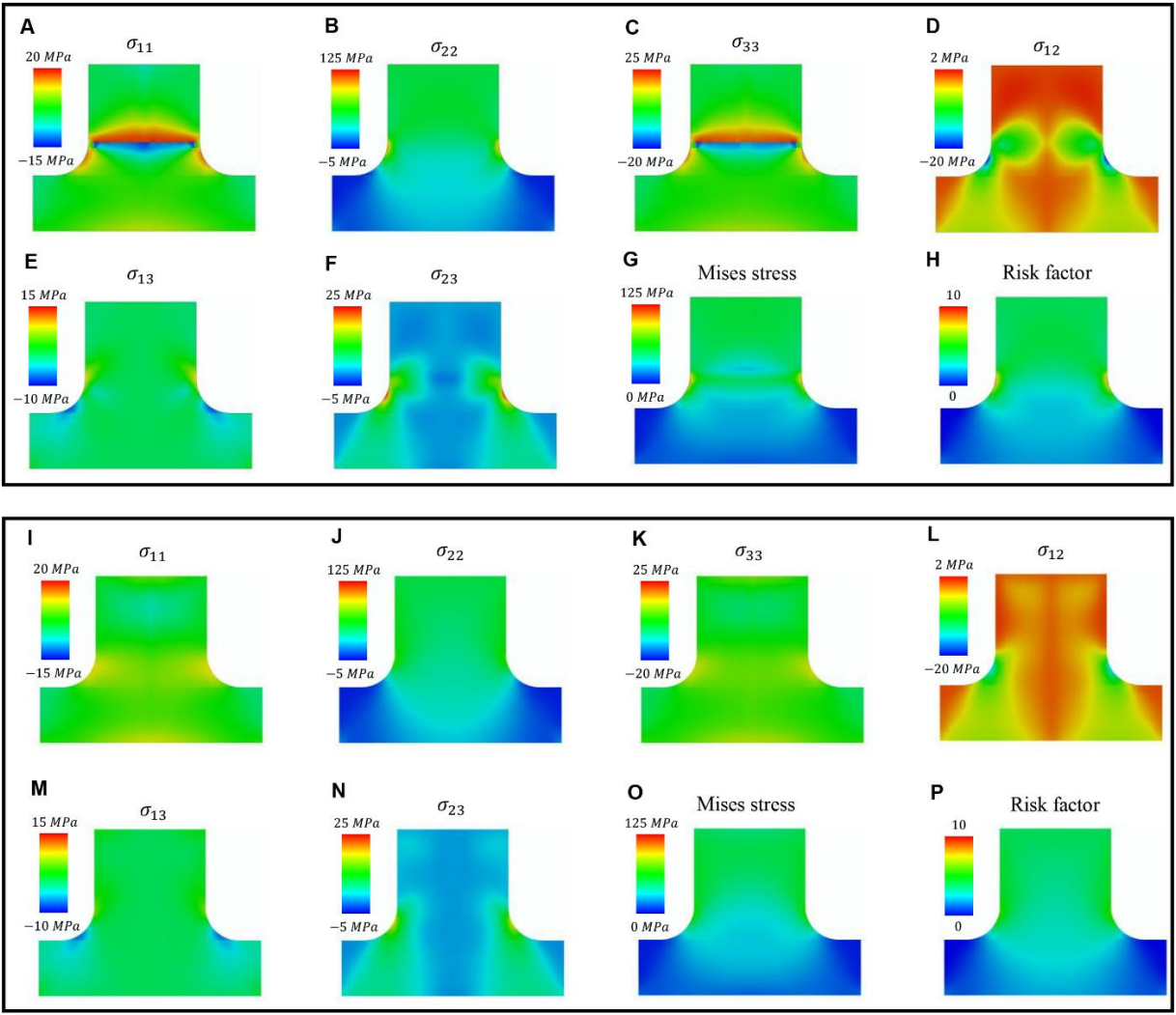

**Figure 3 | Effect of gradient mineralization scale and fibril angular dispersion on stress fields.** **A-H.** Axisymmetric contours of $\sigma_{11}$, $\sigma_{22}$, $\sigma_{33}$, $\sigma_{12}$, $\sigma_{13}$, $\sigma_{23}$, von Mises stress, and the risk factor, respectively, under discontinuous mineralization scale and angular dispersion fields. **I-P.** Corresponding contours under gradient fields. All stress components are shown in the local coordinate system.

With discontinuous mineralization scale and fibril angular dispersion, the normal stress components transverse to the loading direction, $\sigma_{11}$ and $\sigma_{33}$, exhibit pronounced stress concentrations near the discontinuous interface. In particular, $\sigma_{11}$ exhibits tensile stresses above and compressive stresses below the interface, reflecting sharp discontinuities arising from abrupt changes in material anisotropy and elastic modulus. In contrast, gradient fields effectively smooth these transitions, thereby eliminating or substantially reducing stress discontinuities.

The normal stress along the loading direction, $\sigma_{22}$, remains continuous even under discontinuous fields, as required by mechanical equilibrium. However, discontinuous mineralization scale and dispersion induce significant $\sigma_{22}$ concentrations along the curved surface adjacent to the discontinuous interface, whereas gradient configurations markedly mitigate these effects. Consistent with the applied tensile loading, $\sigma_{22}$ exhibits the largest magnitude among all normal stress components.

The shear stress components ($\sigma_{12}$, $\sigma_{13}$, and $\sigma_{23}$) also concentrate near the curved surface under discontinuous conditions but are substantially alleviated when gradients are introduced. Consequently, the von Mises stress exhibits sharp localization at the curved surface in the discontinuous case, whereas such concentrations are largely eliminated under gradient conditions. This behavior arises because the von Mises stress aggregates contributions from individual stress components whose discontinuities are effectively smoothed by gradient transitions.

The risk factor exhibits spatial distributions that closely mirror those of the von Mises stress under both discontinuous and gradient fields. The strong correlation indicates that the proposed risk factor is well-suited for evaluating stress-induced fibril failure risk and can serve as a reliable scalar failure criterion. Importantly, the risk factor not only reproduces the trends of conventional stress measures but also incorporates fibril orientation effects, thereby providing a more physiologically relevant metric for assessing failure susceptibility of collagen fibers. To further assess its robustness, we examine a case involving localized mineralization, an abnormal condition previously associated with smoking and other pathological factors. The results are presented in **Supporting Information (SI)** (Section 3, Fig. S2). These additional simulations confirm that the risk factor effectively captures the salient features of the stress fields under pathological scenarios [34-36], highlighting its applicability to both healthy and diseased entheses.

**Performance of the CNNFP**

In this section, the performance of the CNNFP on both the $\sigma_{22}$ and risk factor fields are evaluated. To assess its predictive capability for $\sigma_{22}$, we randomly select three validation samples and compare the CNNFP-predicted fields with their corresponding ground-truth outputs (Fig. 4A-F). The predicted and true fields are in close agreement, capturing not only the location but also the shape and extent of both high- and low-stress bands and islands. An additional observation is that the quilt-like patterns in the ground-truth fields originate from the specific output mode used in *Abaqus*, which is found to facilitate effective training of CNNFP. Interestingly, the CNNFP results reproduce these patterns in a smoother form. This smoothness, likely stemming from the configuration of the transposed convolutional layers in the network architecture, constitutes a clear advantage, as it suppresses numerical artifacts while preserving the essential spatial characteristics of the stress fields. More importantly, such smooth and stable predictions enhance the CNNFP's suitability for downstream applications, including optimization studies and biomimetic material design, where stable and interpretable field predictions are critical.

To quantify predictive accuracy, we compute the pixel-wise root mean square error (RMSE) between all fields in the validation dataset and their corresponding CNNFP predictions (Fig. 4G). The RMSE distribution is relatively uniform across the domain, with most values below 0.03. Regions of elevated RMSE are localized near the curved surface, which can be attributed to the steep stress gradients in that area.

A similar analysis is performed for the risk factor fields. As shown in Fig. 5A–F, the CNNFP predictions closely match their ground-truth counterparts for three randomly selected validation samples. The predicted fields successfully capture both the distribution and magnitude of high-risk regions. Consistent with the $\sigma_{22}$ results, the overall pixel-wise RMSE remains below 0.03 (Fig. 5G). Compared to $\sigma_{22}$, however, the RMSE distribution for the risk factor appears more discretized. This behavior is expected, as the risk factor depends not only on local stress but also on fibril orientation distributions, which introduce additional variability and complexity into the field.

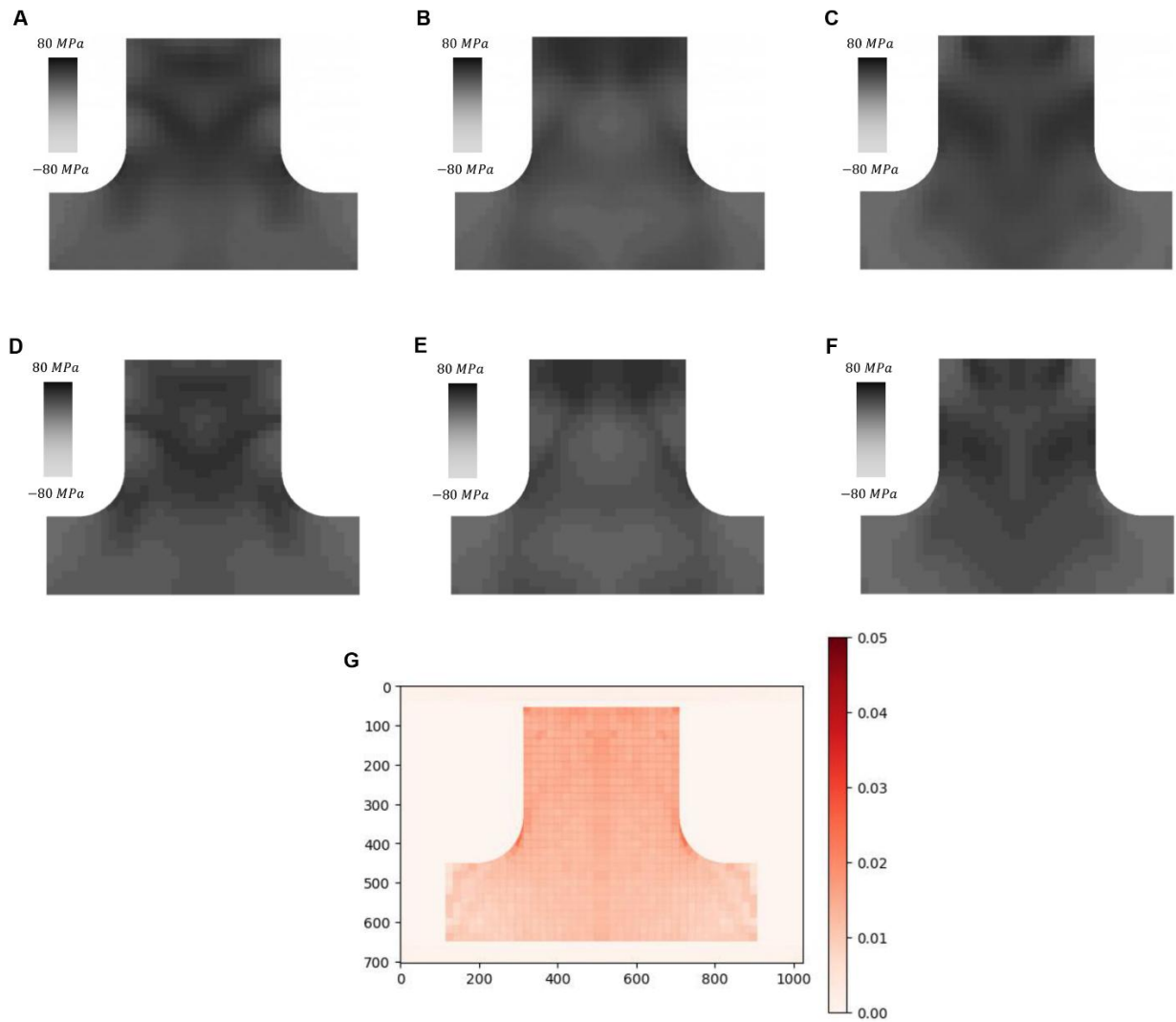

**Figure 4 | Performance of the CNNFP on $\sigma_{22}$ fields. A-C.** CNNFP-predicted $\sigma_{22}$ fields for three randomly selected validation samples. **D-F.** Corresponding ground truth $\sigma_{22}$ fields. **G.** Pixel-wise RMSE between CNNFP predictions and ground truth $\sigma_{22}$ fields across the validation dataset.

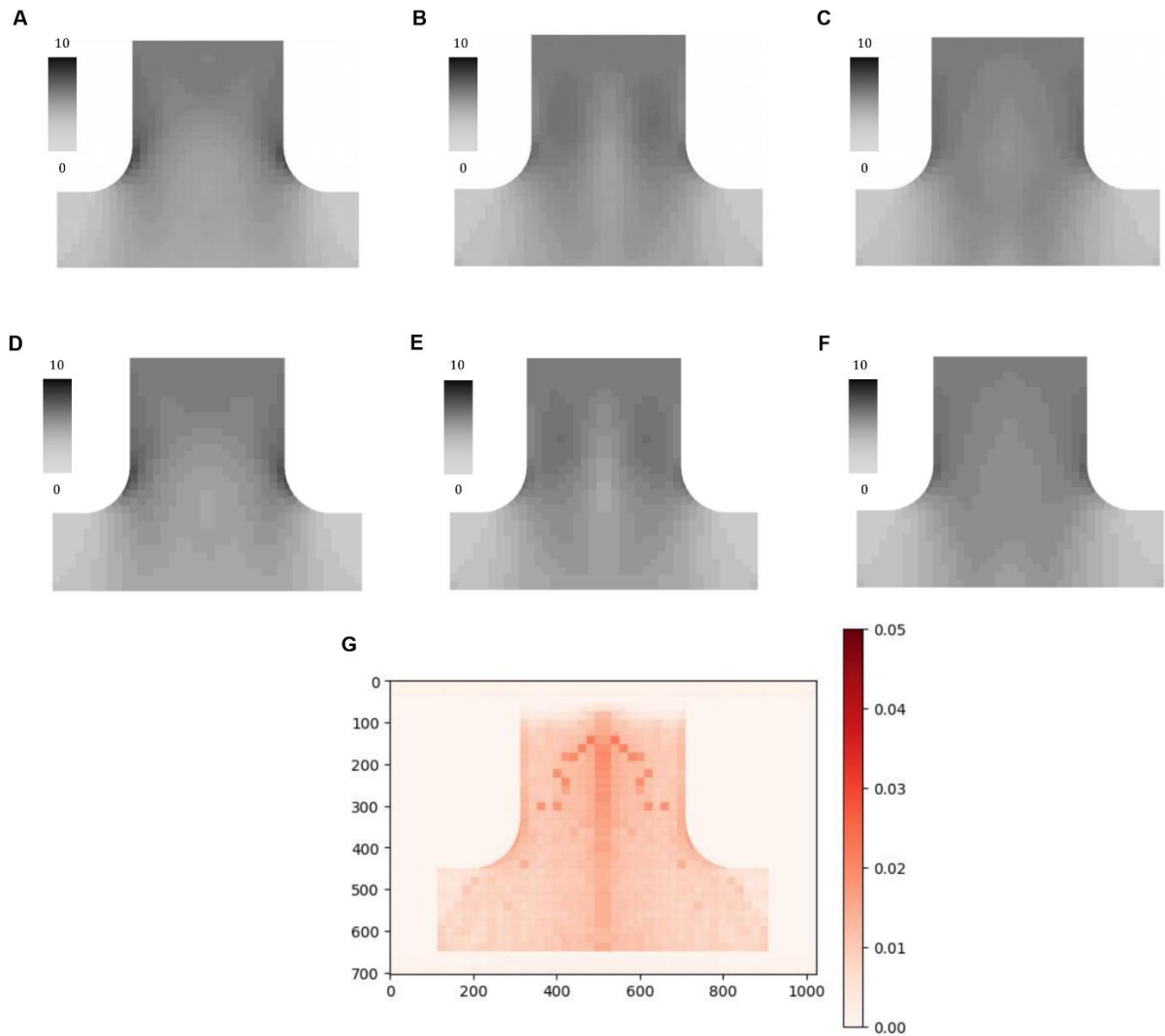

**Figure 5 | Performance of the CNNFP on risk factor fields. A-C.** CNNFP-predicted risk factor fields for three randomly selected validation samples. **D-F.** Corresponding ground-truth risk factor fields. **G.** Pixel-wise RMSE between CNNFP predictions and ground truth risk factor fields across the validation dataset.

Overall, these results demonstrate that the CNNFP achieves high predictive accuracy across both stress and risk factor fields while producing smooth, artifact-suppressed outputs well-suited for downstream analyses. These findings underscore the robustness and strong generalization capability of the framework, reinforcing its potential as a powerful tool for elucidating structure–property relationships and guiding the rational design of bioinspired engineering materials.

**Optimization of the mineralization scale and fibril orientation fields**

This section demonstrates the CNNFP framework's capability for inverse design by directly optimizing spatially varying input fields to achieve a prescribed mechanical objective. Using the field optimization procedure described in the **Methods** Section, we formulate an optimization problem aimed at minimizing the maximum value of the fiber-level risk factor field. To ensure physical relevance, all five input fields—mineralization scale, angular dispersion, and the $x'$, $y'$, and $z'$ components of the mean fibril orientation—are constrained to remain within their prescribed physiological ranges throughout the optimization process.

Starting from randomly generated input fields, as shown in Fig. S4 in the **SI**, the resulting risk factor field exhibits pronounced localization near the curved surface (Fig. 6A), consistent with the stress concentration patterns observed in earlier analyses. In contrast, the optimized risk factor field predicted by the CNNFP (Fig. 6B) displays a markedly more uniform distribution, successfully achieving the objective of suppressing peak failure risk. Notably, the elevated risk region typically observed near the curved surface is reduced to near-baseline levels, indicating a fundamental redistribution of load-bearing mechanisms through coordinated changes in mineralization and fibril orientation.

To validate the optimization results, the optimized input fields were subsequently evaluated using FEM simulations in *Abaqus*, yielding the risk factor field shown in Fig. 6C. The close agreement between the CNNFP-predicted field (Fig. 6B) and the ground-truth result (Fig. 6C) confirms the accuracy of the field predictor and the effectiveness of the optimization strategy.

Fig. 6D-H present the optimized input fields, including the mineralization scale distribution, angular dispersion, and the $x'$, $y'$, and $z'$ components of the mean fibril orientation. A key and nontrivial outcome is that mineralization scale gradients in the transverse direction emerge as an important design feature. Rather than remaining spatially uniform, the mineralization scale degree is reduced in the core region, increases radially outward, and then decreases toward the periphery. Along the longitudinal axis, mineralization scale is low on the tendon side, increases sharply, and subsequently plateaus toward the bone side—an overall trend consistent with experimental observations of tendon–bone enthesis. The optimized angular dispersion field exhibits a similar longitudinal trend, increasing from the tendon side to the bone side in the core region, while remaining relatively low near the curved surface. This configuration indicates that highly aligned fibrils are mechanically advantageous in the vicinity of the curved interface, where the principal

stress direction rotates gradually, and maintaining fibril alignment in this region enables more efficient load transfer.

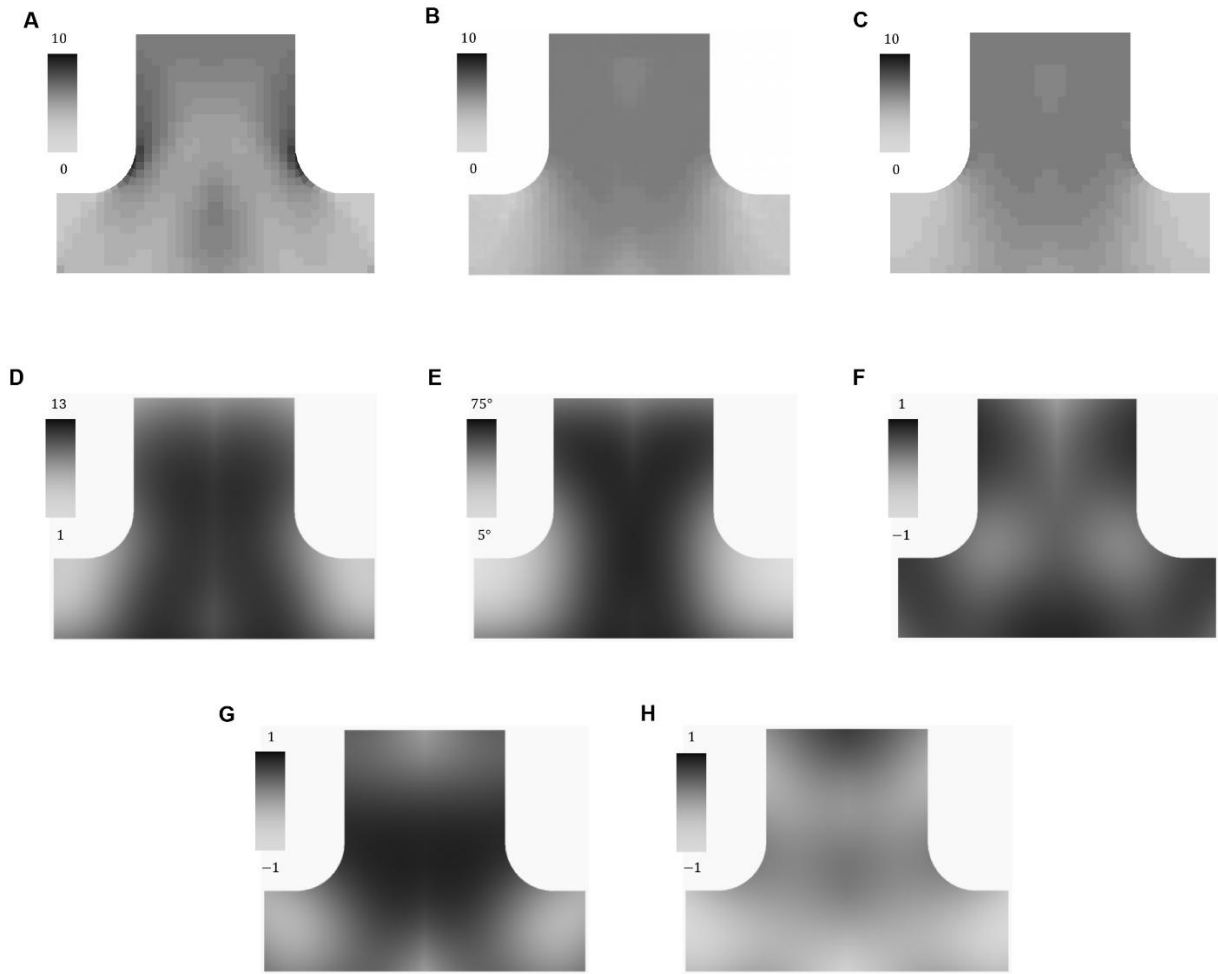

**Figure 6 | Optimization of the mineralization scale and fibril orientation fields. A.** Predicted risk factor field corresponding to randomly generated input fields. **B.** Risk factor field predicted by the CNNFP based on the optimized input fields. **C.** Ground-truth risk factor field obtained from Abaqus using the same optimized input fields. **D-H.** Optimized input fields of mineralization scale, angular dispersion, and the $x'$, $y'$, and $z'$ components of mean fiber orientation.

The optimized mean fibril orientation fields further reinforce this interpretation. Within the tendon region, fibrils are primarily aligned along the longitudinal direction, as evidenced by the dominance of the $y'$ component. In contrast, near the curved surface, the $x'$ and $z'$ components remain approximately constant while the $y'$ component decreases smoothly along the curve. This behavior reflects a gradual and continuous reorientation of fibrils that preserves alignment parallel to the curved interface, thereby minimizing stress concentration and enhancing mechanical robustness.

Overall, these results demonstrate that optimized configurations of mineralization and fibril orientation can effectively suppress localized stress concentrations, particularly near curved surfaces, and promote a more uniform distribution of failure risk. Beyond improved mechanical performance, the optimized field patterns provide valuable insights into the underlying structure–property relationships and the mechanics governing load transfer at heterogeneous interfaces. Specifically, the emergence of coordinated longitudinal and transverse gradients, coupled with locally aligned fibril architectures, reveals fundamental design principles for mitigating material anisotropy and stiffness mismatches.

## Discussion

This study elucidates the mechanics of the tendon–bone enthesis at the level of individual interface fibers, with a focus on the coupled effects of mineralization, fibril angular dispersion, and mean fibril orientation. By developing an FEM of the interface fiber-bone enthesis that explicitly incorporates these spatially varying fields, we resolve local stress distributions and introduce a risk factor that integrates both stress state and fibril orientation. This scalar metric provides a physiologically relevant measure of local failure susceptibility, enabling systematic and quantitative comparisons across different structural configurations.

Our results highlight the mechanical significance of graded structural organization at soft–hard tissue interfaces. Comparisons between gradient and discontinuous configurations reveal that smooth transitions in mineralization scale and fibril dispersion can substantially mitigate or even eliminate stress concentrations and elevated failure risk that otherwise arise at sharp interfaces between mechanically dissimilar materials. These results underscore the critical role of spatially graded composition and microstructure in preserving the mechanical resilience of tendon–bone attachment.

To efficiently explore these high-dimensional structure–property relationships, we develop and validate a CNNFP framework. The CNNFP accurately reproduces stress and risk factor fields while providing orders-of-magnitude improvements in computational efficiency relative to direct

FEM simulations. Notably, the CNNFP generates smooth, artifact-suppressed predictions that enhance interpretability and robustness for downstream analyses. Leveraging this capability, we further establish an optimization framework that identifies field configurations minimizing peak failure risk. The optimized solutions reveal nonintuitive yet mechanically meaningful design principles, including the importance of transverse mineralization scale gradients and the mechanical advantage of maintaining highly aligned fibrils near curved interfaces. These features collectively promote more uniform stress distributions and suppress localized failure.

Beyond advancing fundamental understanding of enthesis mechanics, the optimized field patterns offer translatable insights into underlying structure–property relationships and governing mechanical rules for heterogeneous interfaces. Importantly, these insights provide a mechanistic foundation for translating enthesis-inspired strategies into the rational design of engineered graded composites and architected materials with enhanced mechanical resilience, damage tolerance, and functional robustness. With recent advances in fabrication and synthesis techniques, it has become increasingly feasible to engineer materials with spatially programmed composition and microstructure that emulate nature's functionally graded architectures [37-39]. The methods and findings presented here are therefore both transferable and scalable, offering actionable design principles for biomimetic, heterogeneous synthetic materials capable of unprecedented performance. Potential applications span lightweight and resource-efficient structural systems, soft and humanoid robotics, biomechanically compatible prosthetics, and mechanically robust energy-storage and protective components.

Taken together, the integrated modeling, ML, and optimization framework introduced in this work provides both mechanistic insight into natural tendon–bone enthesis and a generalizable pathway for designing synthetic graded materials with tailored, high-performance mechanical behavior. Looking forward, this framework can be further enhanced by incorporating advanced optimization strategies, such as active learning and deep reinforcement learning [40-42], to more efficiently explore high-dimensional design spaces and competing design solutions. Coupling these approaches with targeted experimental validation and emerging fabrication methods, including additive manufacturing and spatially programmable material synthesis [43-45], will further accelerate translation from biological inspiration to engineered realization.

# Methods

## Spatial description of mineralization and fibril orientation

Mineralization and fibril orientation are inherently spatially varying within the tendon–bone enthesis. To quantify the local degree of mineralization, we define a field variable, referred to as the mineralization scale $m(x,y,z)$, as the multiplicative gain in fibril stiffness induced by mineralization relative to unmineralized fibrils. To characterize the spatial variation of fibril orientation, two field variables are introduced: the mean fibril orientation, $\vec{o}(x,y,z)$, and the angular dispersion, $a(x,y,z)$. The former is a vector field that specifies the mean fibril orientation at each point, whereas the latter is a scalar field that quantifies the local degree of fibril alignment or orientation dispersion.

In the FEM, a global Cartesian coordinate system is first designed, as shown in Fig. 1C. Each element in the model is then assigned to a local orthonormal coordinate system $(o', x', y', z')$, with the origin $o'$ located at the geometric center of the element. The local $y'$-axis is aligned with the global $y$-axis, whereas the $z'$-axis points in the radial direction, as illustrated in Fig. 1C. For convenience, the mean fibril orientation field is expressed in terms of its components in the local coordinate system, denoted as $o1(x,y,z)$, $o2(x,y,z)$, and $o3(x,y,z)$, corresponding to the $x'$, $y'$, and $z'$ directions, respectively. The mean fibril orientation vector field can therefore be written as $\vec{o}(x,y,z) = o1(x,y,z)\vec{i'} + o2(x,y,z)\vec{j'} + o3(x,y,z)\vec{k'}$.

At an arbitrary location $(x,y,z)$ within the model domain, the probability density function $f(x,y,z,\vec{s})$, which describes the likelihood that a collagen fibril at that location is aligned along the direction $\vec{s}$, is defined by Eq. (1) [6,27]:

$$f(x,y,z,\vec{s}) = \frac{e^{\left(-\frac{\phi^2}{2a^2(x,y,z)}\right)}}{\sqrt{2\pi}\,a(x,y,z)\,e^{\left(-\frac{a^2(x,y,z)}{2}\right)} I(x,y,z)} \tag{1}$$

where $\phi$ is the acute angle between $\vec{s}$ and $\vec{o}(x,y,z)$ (as illustrated in Fig. 1C), $a(x,y,z)$ is the angular dispersion, and $I(x,y,z)$ is the probability complementary coefficient, defined by Eq. (2):

$$I(x,y,z) = \frac{1}{2}\int_{\Omega} \frac{e^{\left(-\frac{\phi^2}{2a^2(x,y,z)}\right)}}{\sqrt{2\pi}\,a(x,y,z)\,e^{\left(-\frac{a^2(x,y,z)}{2}\right)}}\, d\vec{s} \tag{2}$$

where $\Omega$ is a unit sphere centered at $(x, y, z)$, and $d\vec{s}$ is the differential area vector on $\Omega$ pointing outward.

In essence, Eq. (1) prescribes an axisymmetric distribution of fibril orientations about the local mean fibril orientation $\vec{o}(x, y, z)$, with the spread of orientations governed by the local angular dispersion field $a(x, y, z)$.

**Collective stiffness tensor dependent on mineralization and fibril orientations**

The constitutive behavior of the interface fiber–bone system is developed using a bottom-up multiscale homogenization framework [27]. Effective stiffness tensors of collagen fibrils are derived from their molecular and nano-structural constituents using matrix–inclusion models [27,46,47], with mineralization incorporated via an equivalent stiffness scaling through the mineralization scale $m(x, y, z)$, which accounts for both intra- and extra-fibrillar hydroxyapatite. At each spatial location, the collective stiffness tensor is obtained by averaging the orientation-dependent fibril stiffness tensor over the local fibril orientation distribution, governed by the mean fibril orientation field and the angular dispersion field. This approach yields a spatially varying, anisotropic constitutive model that directly links mineralization and fibril organization to the mechanical response of the interface fiber–bone system. Detailed formulations are provided in Section 1 of the **SI**, where Table S1 lists volume fractions of constitutive components used in this study. The constitutive model is subsequently implemented in the 3D FEM shown in Fig. 1B using the commercial finite element software *Abaqus 2023*.

**Assessment of failure risk using the risk factor**

To quantify failure susceptibility at the material-point level, we define a risk factor $R$ that measures the likelihood of fibril-level failure under a given local stress state. Specifically, at a material point $(x, y, z)$, the risk factor $R$ is determined by Eq. (3):

$$R = max\{\frac{\sigma_n(\sigma,\vec{n})}{\sigma_{f,n}(m,\vec{o},a,\vec{n})} | \vec{n} \in \Omega\} \qquad (3)$$

where $\Omega$ denotes the unit sphere centered at $(x, y, z)$, $\sigma_n$ is the tensile stress in the direction $\vec{n}$, and $\sigma_{f,n}$ is the tensile strength of the collagen fibril network in that direction.

Specifically, $\sigma_n$ is calculated as:

$$\sigma_n = \vec{n} \cdot \boldsymbol{\sigma} \cdot \vec{n} \tag{4}$$

where $\boldsymbol{\sigma}$ is the local Cauchy stress tensor.

The statistic tensile strength of the fibril network, $\sigma_{f,n}$, is defined as:

$$\sigma_{f,n} = \sigma_f(m) N_f f(x,y,z,\phi) S_f + \sigma_m \tag{5}$$

where $\sigma_f(m)$ is the tensile strength of individual collagen fibrils with a mineralization scale of $m$, $N_f$ denotes the number of collagen fibrils contained within a spherical volume $V_s$ whose diameter is equal to the collagen fibril length, $f(x,y,z,\phi)$ is the orientation probability density function evaluated at angle $\phi$, defined as the acute angle between $\vec{n}$ and mean fibril orientation $\vec{o}$, $S_f$ is the cross-sectional area of the collagen fibrils, and $\sigma_m$ represents the strength of the surrounding matrix. Previous experimental and computational studies have shown that the tensile strength of collagen fibrils, $\sigma_f$, exhibits a non-monotonic dependence on mineralization, initially increasing and then decreasing with increasing mineral content [5,19,25]. To establish a quantitative relationship between $\sigma_f$ and $m$, we perform molecular dynamics simulations based on a previously developed coarse-grained model of mineralized collagen fibrils [21]. Details of the molecular simulations and other collagen fibril-related parameters are provided in Section 2 (Fig. S1, Table S2, and Table S3) of the **SI**.

**CNNFP-enabled field prediction and optimization framework**

Based on the developed continuum model of tendon-bone interface fibers, we formulate a generalized problem in which a set of five spatially varying input fields governs the mechanical response of the model and produces one or more output fields of interest. Specifically, five input fields include the mineralization scale field $m(x,y,z)$, the angular dispersion field $a(x,x,z)$, and the three components of the mean fibril orientation field, $o1(x,y,z)$, $o2(x,y,z)$, and $o3(x,y,z)$, corresponding to the local $x'$, $y'$, and $z'$ directions, respectively. To simplify the problem, we enforce axisymmetry about the $y$-axis for all five input fields. Therefore, the input fields $m$, $a$, $o1$, $o2$, and $o3$ depend only on the axial coordinate $y$ and the radial distance from the $y$-axis, defined as $r = \sqrt{x^2 + z^2}$. We further introduce a unified notation $f^{(l)}(r,y)$ for $l = 1,2,3,4,5$, such that:

$$f^{(1)}(r,y) = m(x,y,z), \tag{6}$$

$$f^{(2)}(r,y) = a(x,y,z), \qquad (7)$$

$$f^{(3)}(r,y) = o1(x,y,z), \qquad (8)$$

$$f^{(4)}(r,y) = o2(x,y,z), \qquad (9)$$

$$f^{(5)}(r,y) = o3(x,y,z). \qquad (10)$$

The output fields of the continuum model include, but are not limited to, stress or strain components, risk factor $R$ field, and other mechanical quantities of interest. This mapping from the five inputs to the output field is obtained from numerical simulations conducted using *Abaqus 2023*. With hardware available in this study (Intel® Core™ i9-10900 CPU, 10 cores, 20 threads, 2.8 GHz base frequency), a single forward simulation requires approximately 3 minutes of wall-clock time, which enables the generation of a sufficiently large training dataset within a reasonable computational budget.

The CNNFP is constructed based on a convolutional neural network architecture [48,49], as shown in Fig. 7A. We train the CNNFP with a training dataset comprising 1600 samples and an additional 400 samples for validation. Each sample consists of six grayscale images, in which five represent the input fields: mineralization, angular dispersion, and the $x'$, $y'$, and $z'$ components of the mean fibril orientation, and one corresponds output field. All fields are represented on a $704 \times 1024$ grid.

The output of CNNFP can be any field of interest, such as a stress component field or a risk factor field. As a representative example, we select $\sigma_{22}$ and the risk factor to demonstrate the performance of the CNNFP, as they effectively characterize the mechanical state and failure susceptibility of our model problem. The adaptive moment estimation optimizer is used in the training process, with the mean squared error used as the loss function. After approximately 300 epochs, both training and validation errors for both $\sigma_{22}$ and the risk factor converges to values below or on the order of $10^{-4}$, indicating accurate and stable learning. The training error versus epoch plot is shown in Fig. S3 of the **SI**. Once trained, the CNNFP enables rapid and accurate field prediction, making it well-suited for downstream optimization and design applications.

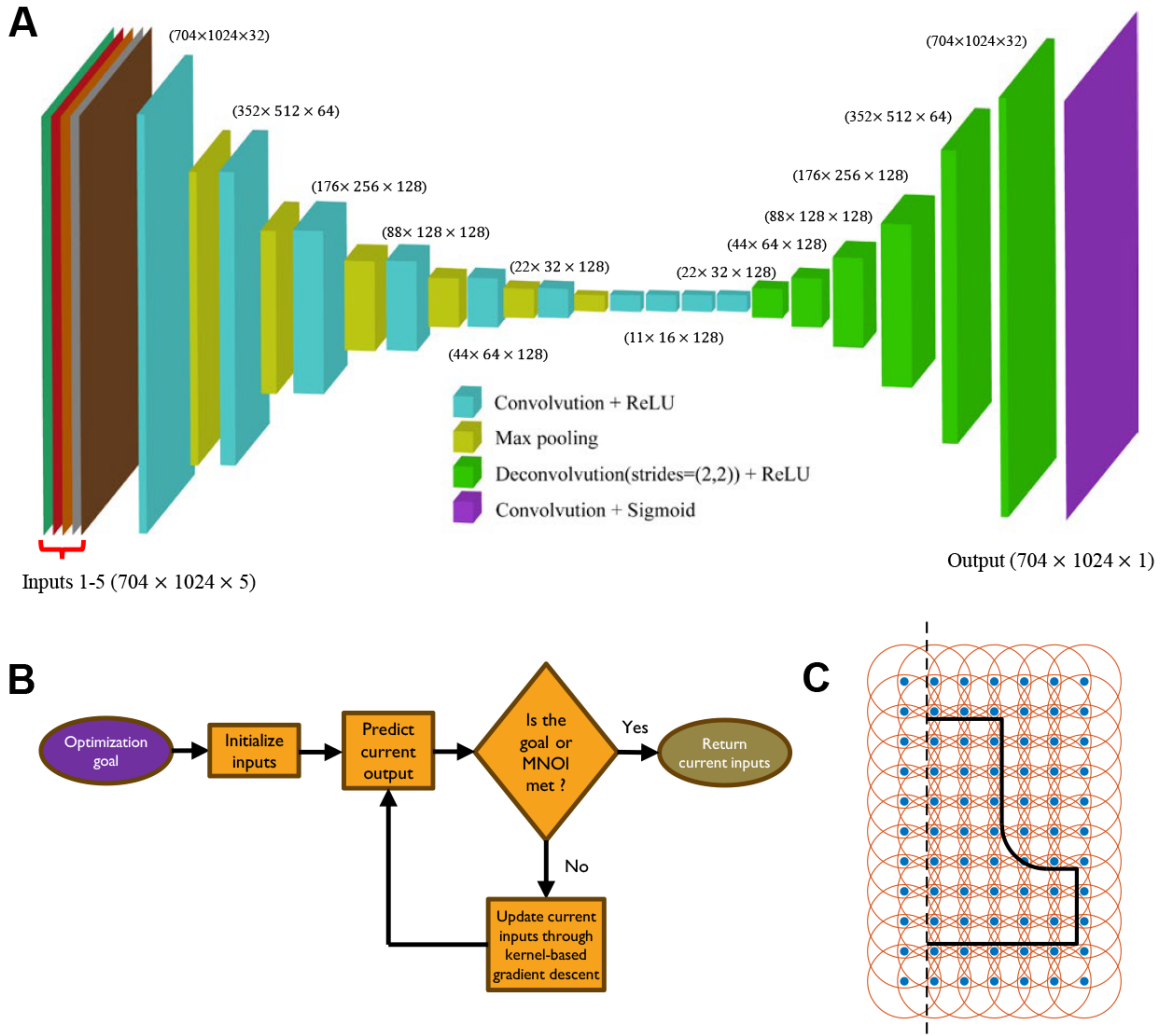

**Figure 7 | CNNFP-enabled field prediction and optimization framework. A.** Schematic of the CNNFP predicting target output fields from five input fields. **B.** Workflow of the field optimization procedure enabled by the trained CNNFP. **C.** Illustration of the kernel-based field generator, in which values at kernel nodes are iteratively updated using a gradient descent algorithm to produce continuous input fields.

**Procedure for performing field optimization using the trained CNNFP**

Leveraging the trained CNNFP, we implement an efficient procedure for optimizing spatial input fields to achieve prescribed mechanical objectives, as illustrated in Fig. 7B. The procedure begins by specifying an optimization target, for instance, constraining the risk factor field below a desired threshold. An initial set of five input fields is then generated and passed through the CNNFP to obtain the corresponding output field, and the discrepancy between the predicted output and the target is quantified using a loss function. Using a kernel-based gradient descent strategy, the gradient of the loss function with respect to the kernel nodal values associated with the five input fields is computed. These kernel nodal values of five input fields are then updated simultaneously in the negative gradient direction with a prescribed step size to minimize the loss. If the optimization objective is satisfied, the current input fields to the CNNFP are accepted as the optimal solution. Otherwise, the input fields are iteratively updated until either the loss drops below the threshold or the maximum number of iterations is reached. More details about this process in included in Section 6 of the **SI**.

For convenience and consistency, all spatial input fields are first normalized to the interval [0, 1] before optimization. The corresponding physical upper and lower bounds of each input field are provided in Table S4 of the **SI**. After normalization, the fields are parameterized using a kernel-based representation defined by a finite set of kernel nodes (Fig. 7C), whose amplitudes are treated as the optimization variables during the iterative process. This parameterization step substantially reduces the dimensionality of the optimization problem while enforcing spatial smoothness and physical plausibility of the generated fields.

Specifically, each normalized field $\bar{f}^{(l)}(r,y)$ is expressed as a weighted superposition of kernel functions centered at prescribed node locations:

$$\bar{f}^{(l)}(r,y) = \sum_{i=1}^{n} \alpha_i^{(l)} k\big((r,y),(r_i,y_i)\big) \quad (l = 1,2,3,4,5) \quad (11)$$

where $\alpha_i^{(l)}$ denotes the kernel nodal value associated with field $l$ and kernel node $i$, $(r_i, y_i)$ specifies the spatial location of kernel node $i$, and $n$ is the total number of kernel nodes. The number and locations of kernel nodes are held constant to ensure consistent parameterization during optimization. A detailed list of kernel node locations is provided in Table S5 of the **SI**.

The kernel function $k((r,y),(r_i,y_i))$ is chosen as a Gaussian radial basis function:

$$k\big((r,y),(r_i,y_i)\big) = \exp\left(-\frac{(r-r_i)^2+(y-y_i)^2}{2\sigma^2}\right) \quad (12)$$

where $\sigma$ controls the spatial extent of each kernel and therefore the smoothness of the resulting field. Larger values of $\sigma$ produce smoother, more slowly varying fields, whereas smaller values allow for finer spatial variation.

During optimization, a gradient descent algorithm essentially updates the kernel nodal values $\alpha_i^{(l)}$ while keeping the node locations fixed. Through this process, the input fields remain smooth, interpretable, and compatible with the underlying continuum assumptions of the mechanical model, avoiding nonphysical oscillations or discontinuities that could arise from direct pixel-wise optimization.

This kernel-based, CNNFP-enabled optimization framework constitutes an important methodological advance, enabling efficient and physically consistent inverse design of spatially heterogeneous materials directly at the microstructure level, an approach that is difficult to achieve with conventional finite element–based optimization strategies.

**Data Availability**

The authors confirm that the data supporting the findings of this study are available within the article and the supplemental materials. The models and codes for the presented data are available at https://doi.org/10.5281/zenodo.18307669.

**Acknowledgments**

The authors acknowledge funding support from SC TRIMH (P20GM121342) and the New Innovator in Food & Agriculture Research Award from the Foundation for Food & Agriculture Research (23-000603). Clemson University is acknowledged for the generous allotment of computational time on the Palmetto cluster.

**Author Contributions**

Z.M. ideated and supervised the research. Z.Y. performed the simulations and data analysis. All authors contributed to the writing. All authors have read and approved of the final manuscript.

**Competing Interests**

All authors declare no financial or non-financial competing interests.

# *Support Information for*

# Machine Learning-Enabled Mechanical Analysis and Optimization of Bioinspired Functionally Graded Materials

Zhangke Yang[1] and Zhaoxu Meng[1*]

[1] Department of Mechanical Engineering, Clemson University, Clemson, SC, 29634, USA

Corresponding author:

* zmeng@clemson.edu

# Table of Contents

# 1. Collective Stiffness Tensor Dependent on the Mineralization Scale and Orientation of Fibrils

The constitutive model employed in this study is developed using a bottom-up multiscale approach. It begins by deriving the effective mechanical properties of collagen fibrils from their fundamental constituents—collagen, water, non-collagenous proteins, and hydroxyapatite crystals—by treating the system as a sequence of matrix-inclusion problems [1,2]. At each material point, the local constitutive behavior is then determined by homogenizing the properties of collagen fibrils according to their orientation distribution, thereby representing the collective mechanical response of a fibril network at a larger length scale.

Collagen fibril formation originates from collagen molecules assembling through overlap and crosslinking, resulting in so-called wet collagen. Water and non-collagenous proteins occupy the intermolecular spaces within this structure. Accordingly, the elastic properties—specifically, the effective stiffness tensor—of wet collagen are modeled by treating collagen molecules as the matrix and the water–protein mixture as inclusions. Assuming the inclusions to be approximately cylindrical in shape, the effective stiffness tensor of wet collagen, $\boldsymbol{C}_{wc}$, is calculated using the matrix–inclusion formulation given by Eq. (S1) [1,3,4]:

$$\boldsymbol{C}_{wc} = (1 - f_{wp})\boldsymbol{C}_{col} + f_{wp}\boldsymbol{C}_{wp}\left[\boldsymbol{I} + \boldsymbol{P}_{wp}^{col}(\boldsymbol{C}_{wp} - \boldsymbol{C}_{col})\right]^{-1}:$$

$$\left\{(1 - f_{wp})\boldsymbol{I} + f_{wp}\left[\boldsymbol{I} + \boldsymbol{P}_{wp}^{col}:(\boldsymbol{C}_{wp} - \boldsymbol{C}_{col})\right]^{-1}\right\}^{-1} \tag{S1}$$

where $f_{wp}$ is the volume fraction of the water-protein mixture, $\boldsymbol{C}_{col}$ is the stiffness tensor of collagen, $\boldsymbol{C}_{wp}$ is the stiffness tensor of the water-protein mixture, $\boldsymbol{I}$ is the fourth-order identity tensor, and $\boldsymbol{P}_{wp}^{col}$ is the Hill tensor of the water-protein mixture embedded in the collagen matrix, whose components can be calculated based on the shape of the inclusion and on the stiffness tensor of the surrounding matrix [1].

The mineralization scale $m$ corresponds to the volume fraction of hydroxyapatite. Following prior studies, hydroxyapatite crystals are assumed to be interpenetrated by water and non-collagenous proteins [5]. Accordingly, a hydroxyapatite foam is introduced to represent the composite of hydroxyapatite, water, and non-collagenous proteins [1]. The effective stiffness tensor of the hydroxyapatite foam, $C_{hf}$, is computed using the self-consistent homogenization scheme given by Eq. (S2) [1]:

$$C_{hf} = \Sigma_{r\in\{ha,wp\}} f_r C_r : [I + P_r^0 : (C_r - C_{hf})]^{-1} :$$

$$\{\Sigma_{s\in\{ha,wp\}} f_s [I + P_s^0 : (C_s - C_{hf})]^{-1}\}^{-1} \quad (S2)$$

where $ha$ denotes the hydroxyapatite, $wp$ represents the water-protein mixture, $f_{ha}$ and $f_{wp}$ are the corresponding volume fractions, $C_{ha}$ and $C_{wp}$ are the phase stiffness tensors, $P_{ha}^0$ and $P_{wp}^0$ are the corresponding Hill tensors.

Collagen fibrils are subsequently modeled as a composite system consisting of wet collagen and hydroxyapatite foam, with the assumption that needle-shaped hydroxyapatite foam is embedded within the wet collagen matrix, aligning along the main axis of the collagen fibril [6]. Given the effective stiffness tensor of wet collagen and hydroxyapatite foam, the effective stiffness tensor of collagen fibril, $C_{fbl}$, is predicted as Eq. (S3) [1].

$$C_{fbl} = (1 - f_{hf})C_{wc} + f_{hf}C_{hf} : [I + P_{hf}^{wc} : (C_{hf} - C_{wc})]^{-1} :$$

$$\{(1 - f_{hf})I + f_{hf}[I + P_{hf}^{wc} : (C_{hf} - C_{wc})]^{-1}\}^{-1} \quad (S3)$$

where $f_{hf}$ is the volume fraction of hydroxyapatite foam, and $P_{hf}^{wc}$ can be determined from the Hill tensor corresponding to a cylindrical inclusion embedded in a transversely isotropic medium [7].

The volume fraction parameters used to calculate the effective stiffness tensor of the unmineralized collagen fibril, $C_{fbl}^0$, are summarized in Table S1.

**Table S1**. Volume fractions of constitutive components in unmineralized collagen fibrils.

| Equation | $f_{wp}$ | $f_{ha}$ | $f_{hf}$ |
|---|---|---|---|
| Eq. (S1) | 0.35 | - | - |
| Eq. (S2) | 1.00 | 0.00 | - |
| Eq. (S3) | - | - | 0.43 |

Based on the definition of mineralization scale, the equivalent stiffness tensor of a collagen fibril with a mineralization scale of $m$, $\widetilde{C}_{fbl}(m)$, is related to $C_{fbl}^0$ through Eq. (S4).

$$\widetilde{C}_{fbl}(m) = m C_{fbl}^0 \tag{S4}$$

At each spatial location within the tendon–bone enthesis, fibrils are oriented in different directions. To determine the collective material properties at a given point, the equivalent stiffness tensor of individual fibrils is homogenized over the local fibril orientation distribution. The resulting collective stiffness tensor, $C_{clt}(x, y, z)$, is computed as:

$$C_{clt}(x, y, z) = \int_\Omega (R(\vec{s}) \cdot \widetilde{C}_{fbl} \cdot R^T(\vec{s})) f(x, y, z, \vec{s}) \, d\vec{s} \tag{S5}$$

where $C_{clt}$ is represented in the local coordinate system whose z-axis points to $\vec{o}$, i.e., the mean fibril orientation, $\Omega$ is a unit sphere centered at $(x, y, z)$, $R(\vec{s})$ is the rotation matrix that transforms a coordinate system with z-axis pointing to $\vec{s}$ into another coordinate system with z-axis pointing to $\vec{o}$, $f(x, y, z, \vec{s})$ is the probability density function for finding a fibril aligned in $\vec{s}$, and $d\vec{s}$ is the differential area vector on $\Omega$, pointing outward.

This homogenization procedure yields a spatially varying, anisotropic stiffness tensor that directly links fibril-level mineralization and orientation to the continuum mechanical response of the interface fiber–bone system.

## 2. Trend of the Tensile Strength of Collagen Fibrils $\sigma_f$ with Mineralization Scale $m$

To quantify the dependence of collagen fibril tensile strength, $\sigma_f$, on mineralization scale $m$, we employ a coarse-grained (CG) molecular dynamics (MD) model of collagen fibrils adapted from a previous study [8], as illustrated in Fig. S1A. In this model, tropocollagen (TC) triple helices are represented by magenta beads, while hydroxyapatite (HA) crystals are represented by blue beads. Interactions among beads are governed by a combination of bonded, angular, and nonbonded multi-body potentials. MD simulations are performed by the LAMMPS molecular dynamics package [9].

The total potential energy of the CG collagen fibril system is expressed as:

$$E = E_{Bond} + E_{Angle} + E_{Pair} \tag{S6}$$

where $E_{Bond}$ accounts for bonded interactions between adjacent TC beads or adjacent HA beads, $E_{Angle}$ represents angular interactions among three consecutive TC or HA beads, and $E_{Pair}$ includes all nonbonded interactions between bead pairs, namely TC–TC, HA–HA, and TC–HA interactions.

The bonded force between two adjacent TC beads is described by [8]:

$$F_{B,TC}(r) = -d\phi_{B,TC}(r)/dr \tag{S7}$$

where

$$\frac{d\phi_{B,TC}(r)}{dr} = H(r_{break} - r) \begin{cases} k_{B,TC}^{(0)}(r - r_0) & r < r_1 \\ k_{B,TC}^{(1)}(r - \tilde{r}_1) & r \geq r_1 \end{cases} \tag{S8}$$

where $H(a)$ is the Heaviside function, whose value is zero when $a < 0$, and one when $a \geq 0$. $k_{B,TC}^{(0)}$ and $k_{B,TC}^{(1)}$ are the small and large-deformation spring constants, respectively. $r_{break}$, $r_0$, $r_1$, and $\tilde{r}_1$ are bond length parameters. Continuity of the force is enforced by defining

$$\tilde{r}_1 = r_1 - k_{B,TC}^{(0)}/k_{B,TC}^{(1)}(r_1 - r_0) \tag{S9}$$

The bonded interaction between two adjacent HA beads is modeled by the harmonic bond, with spring constant determined from the Young's modulus of HA crystals [8]:

$$\phi_{B,HA}(r) = k_{B,HA}(r - r_0)^2 \tag{S10}$$

where $k_{B,HA}$ is the harmonic bond spring constant, and $r_0$ is the equilibrium bond length.

The angle potential form for three consecutive TC beads is given by

$$\phi_{A,TC}(\varphi) = \frac{1}{2} k_{A,TC}(\varphi - \varphi_{0,TC})^2 \tag{S11}$$

where $k_{A,TC}$ represents the angle stiffness, $\varphi$ is the current angle formed by the three beads, and $\varphi_{0,TC}$ is the equilibrium angle of TC beads.

Similarly, the angle potential form for three consecutive HA beads is given by

$$\phi_{A,HA}(\varphi) = \frac{1}{2} k_{A,HA}(\varphi - \varphi_{0,HA})^2 \tag{S12}$$

where $k_{A,HA}$ represents the angle stiffness, $\varphi$ is the current angle formed by the three beads, and $\varphi_{0,HA}$ is the equilibrium angle of HA beads.

All nonbonded interactions are described by the Lennard-Jones (LJ) 12-6 potentials in Eq. (S13-S15), including TC–TC, HA–HA, and TC–HA interactions:

$$\phi_{TC-TC} = 4\varepsilon_{TC-TC}\left[\left(\frac{\sigma_{TC-TC}}{r}\right)^{12} - \left(\frac{\sigma_{TC-TC}}{r}\right)^{6}\right] \tag{S13}$$

$$\phi_{HA-HA} = 4\varepsilon_{HA-HA}\left[\left(\frac{\sigma_{HA-HA}}{r}\right)^{12} - \left(\frac{\sigma_{HA-HA}}{r}\right)^{6}\right] \tag{S14}$$

$$\phi_{TC-HA} = 4\varepsilon_{TC-HA}\left[\left(\frac{\sigma_{TC-HA}}{r}\right)^{12} - \left(\frac{\sigma_{TC-HA}}{r}\right)^{6}\right] \tag{S15}$$

where $\sigma_m$ denotes the characteristic interaction distance at which the potential equals zero and $\varepsilon_m$ represents the depth of the potential well for interaction type ($m \in \{TC-TC, HA-HA, TC-HA\}$).

The force-field parameters used in this study are summarized in Table S2.

**Table S2**. Summary of force-field parameters used in this study.

| Equation | Parameter | Value | Equation | Parameter | Value |
|---|---|---|---|---|---|
| (S8), (S9) | $r_{break}$ | 10.5 Å | (S11) | $\varphi_{0,TC}$ | 180° |
| | $r_0$ | 7 Å | (S12) | $k_{A,HA}$ | 230.72 kcal/(mol rad²) |
| | $r_1$ | 9.1 Å | | $\varphi_{0,HA}$ | 180° |
| | $\tilde{r}_1$ | 8.92 Å | (S13) | $\varepsilon_{TC-TC}$ | 2.8 kcal/mol |
| | $k_{B,TC}^{(0)}$ | 17.13 kcal/(mol Å²) | | $\sigma_{TC-TC}$ | 15.96 Å |
| | $k_{B,TC}^{(1)}$ | 195.32 kcal/(mol Å²) | (S14) | $\varepsilon_{HA-HA}$ | 2.8 kcal/mol |
| (S10) | $k_{B,HA}$ | 689.8 kcal/(mol Å²) | | $\sigma_{HA-HA}$ | 3.12 Å |
| | $r_0$ | 4.95 Å | (S15) | $\varepsilon_{TC-HA}$ | 25 or 15 kcal/mol* |
| (S11) | $k_{A,TC}$ | 15 kcal/(mol rad²) | | $\sigma_{TC-HA}$ | 7 Å |

* A value of $25 \, kcal/mol$ is used for most nonbonded $TC - HA$ interactions, whereas $15 \, kcal/mol$ is applied to nonbonded interactions between TC molecular ends and HA.

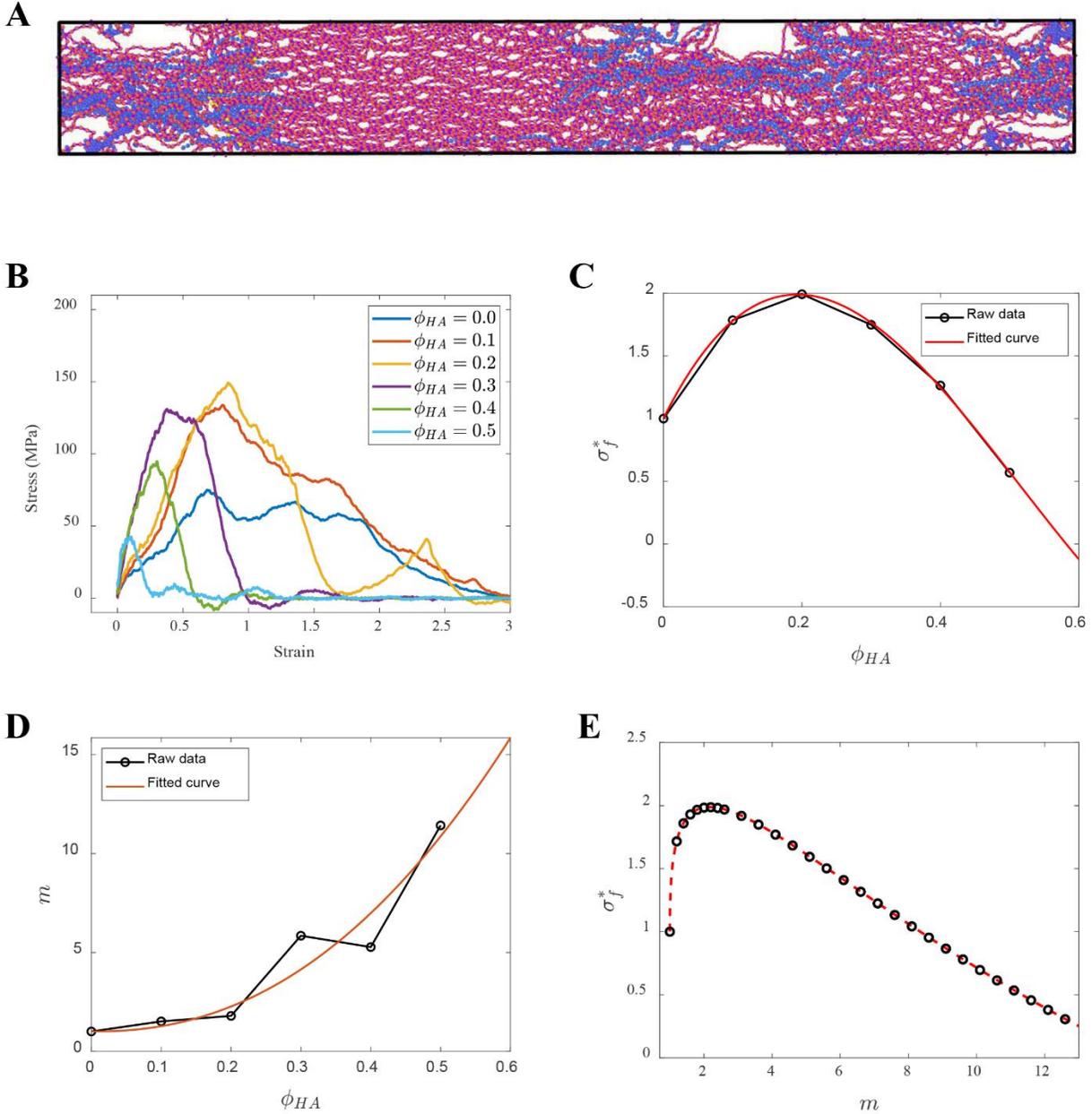

**Figure S1 | CG modeling on $\sigma_f$ of mineralized collagen fibrils. A.** The CG model of the collagen fibril system with a HA volume fraction $\phi_{HA}$ of 0.3. **B.** Stress-strain curves obtained from the tensile tests of CG collagen fibrils with varying HA volume fractions. **C.** $\sigma_f^*$ vs $\phi_{HA}$ and the fitted cubic trend. **D.** $m$ vs $\phi_{HA}$ and the fitted power-law trend. **D.** Relationship between the relative strength $\sigma_f^*$ and the mineralization scale $m$.

The initial structure of the CG collagen fibril model is generated using in-house code according to the prescribed HA volume fraction $\phi_{HA}$. To eliminate possible overlap between generated CG beads, the nonbonded LJ interactions are first switch to a "soft" potential [9], after which the system is energy-minimized for a maximum of 50000 iterations. Initial velocities of the CG beads are then initialized from a Maxwell-Boltzmann distribution at $310\ K$. The system is subsequently equilibrated under the NPT ensemble at $310\ K$ and zero pressure for 50000 time steps, with a time step of $2.66 \times 10^{-3}\ ps$. After equilibration, the nonbonded potentials are switched back to LJ potential using the predetermined parameters. Bond breaking criteria are defined for the TC and HA bonds, with maximum bond lengths of 10.5 Å and 4.998 Å, respectively [8,10]. The system is then re-equilibrated under the NPT ensemble at $310\ K$ and zero pressure for 50000 steps, after which it is ready for subsequent tensile testing. Tensile tests are performed on CG collagen fibril models at a strain rate of $1.88 \times 10^{-3}/ps$, with HA volume fractions between 0 and 0.5. The resulting stress-strain curves are shown in Fig. S1B.

$\sigma_f$ of a collagen fibril s defined as the maximum stress attained in its stress–strain curve. The relative strength, $\sigma_f^*$, is defined as the ratio of $\sigma_f$ to the strength of the unmineralized collagen fibril, $\sigma_{f0}$. The variation of $\sigma_f^*$ with increasing HA volume fraction $\phi_{HA}$ is shown in Fig. S1C, and the data are fitted using a cubic polynomial, yielding the relationship given in Eq. (S16).

$$\sigma_f^*(\phi_{HA}) = 22.83\phi_{HA}^3 - 35.27\phi_{HA}^2 + 11.07\phi_{HA} + 1 \tag{S16}$$

where $0 \leq \phi_{HA} \leq 0.5$.

In addition, the Young's modulus of each collagen fibril is obtained by fitting the quasi-linear slope of the stress–strain curve in the initial deformation regime. The mineralization scale $m$ is calculated as the ratio of the Young's modulus of a mineralized collagen fibril to that of the unmineralized case. The dependence of $m$ on the HA volume fraction $\phi_{HA}$ is shown in Fig. S1D and described by a power-law fit, as given in Eq. (S17).

$$m(\phi_{HA}) = 46.72\phi_{HA}^{2.243} + 1 \tag{S17}$$

where $0 \leq \phi_{HA} \leq 0.5$.

Since both $\sigma_f^*$ and $m$ depend on $\phi_{HA}$, a direct relationship between these two quantities can be established by combining Eq. (S16) and (S17) to eliminate $\phi_{HA}$. This yields the relationship between $\sigma_f^*$ and $m$, as shown in Fig. S1E and expressed in Eq. (S18).

$$\sigma_f^*(m) = 22.83 \left(\frac{m-1}{46.72}\right)^{3/2.243} - 35.27 \left(\frac{m-1}{46.72}\right)^{\frac{2}{2.243}} + 11.07 \left(\frac{m-1}{46.72}\right)^{1/2.243} + 1 \quad \text{(S18)}$$

where $1 \leq m < 11$.

Other collagen fibril-related parameters, including $d$, $L$, $S_f$, $V_f$, $N_f$, $\sigma_{f0}$, and $\sigma_m$, are directly informed from the literature and are summarized in Table S3.

**Table S3**. Summary of collagen fibril-related parameters used in this study.

| Parameter | Symbol | Value used | Literature value |
|---|---|---|---|
| Diameter of collagen fibril | $d$ | $10^{-7}$ m | $\sim 10^{-7}$ m [11,12] |
| Length of collagen fibril | $L$ | $2.828 \times 10^{-6}$ m | $\sim 10^{-6} - 10^{-5}$ m [13-15] |
| Cross-sectional area of collagen fibril | $S_f$ | $7.854 \times 10^{-15}$ m² | $\sim 10^{-14}$ m² (Derived from $d$) |
| Volume of collagen fibril | $V_f$ | $2.221 \times 10^{-20}$ m³ | $\sim 10^{-20} - 10^{-19}$ m³ (Derived from $L$ and $S_f$) |
| Number of collagen fibrils within $V_s$ | $N_f$ | 533 | $\sim 10^2 - 10^4$ (Estimated from $V_s$ and $V_f$) |
| Strength of an unmineralized collagen fibril | $\sigma_{f0}$ | $10^8$ Pa | $\sim 10^8$ Pa [16,17] |
| Strength of the matrix surrounding the collagen fibril | $\sigma_m$ | $10^7$ Pa | $\sim 10^7$ Pa [18,19] |

## 3. Effect of Localized Mineralization on Enthesis Mechanics

In this section, we examine the mechanical consequences of localized mineralization, a pathological feature that has been linked to smoking and other disease-related conditions [20]. Without loss of generality, localized mineralization is modeled as a spherical region of elevated mineral content (approximately 30 $\mu m$ in diameter) centered along the axis of the interface fiber, as illustrated in Fig. S2A. The mineralization scale attains a peak value of 10 at the center of the sphere and gradually decays with increasing radial distance to the baseline level at the outer radius of the spherical region. To isolate the mechanical effects of localized mineralization, we adopt the same gradient angular dispersion (Fig. 2D) and mean fibril orientation along the y-axis (Fig. 2F–H) as in the baseline configuration.

The presence of localized mineralization induces pronounced stress concentrations in all normal stress components ($\sigma_{11}$, $\sigma_{22}$, and $\sigma_{33}$) in and around the mineralized region. Notably, $\sigma_{11}$ and $\sigma_{33}$ exhibit compressive stress concentrations, whereas $\sigma_{22}$ shows amplified tensile stress. This contrast arises from the relative orientations of these stress components relative to the applied axial tensile loading. The transverse components $\sigma_{11}$ and $\sigma_{33}$ are governed primarily by Poisson effect-driven contraction of the surrounding tissue. Because the locally mineralized region is substantially stiffer, it constrains this transverse deformation, leading to a mismatch in lateral strain and the development of compressive stresses in and near the mineralized zone. In contrast, along the loading direction, the stiffer mineralized region experiences smaller axial strains than the adjacent tissue under the same applied load. To satisfy axial strain compatibility, the mineralized region must therefore sustain a higher axial stress, resulting in tensile stress amplification in $\sigma_{22}$.

For the shear stress components ($\sigma_{12}$, $\sigma_{13}$, and $\sigma_{23}$), no pronounced localization is observed within the mineralized core itself. Instead, characteristic belt-like regions of alternating positive and negative shear stresses emerge, extending outward from the mineralized region and oriented at approximately 45° relative to the fiber axis. In addition, significant shear stress concentrations are observed near the curved interface surface, particularly for $\sigma_{12}$ and $\sigma_{23}$, indicating that localized mineralization not only perturbs internal stress distributions but also amplifies stress concentrations at geometric transition regions.

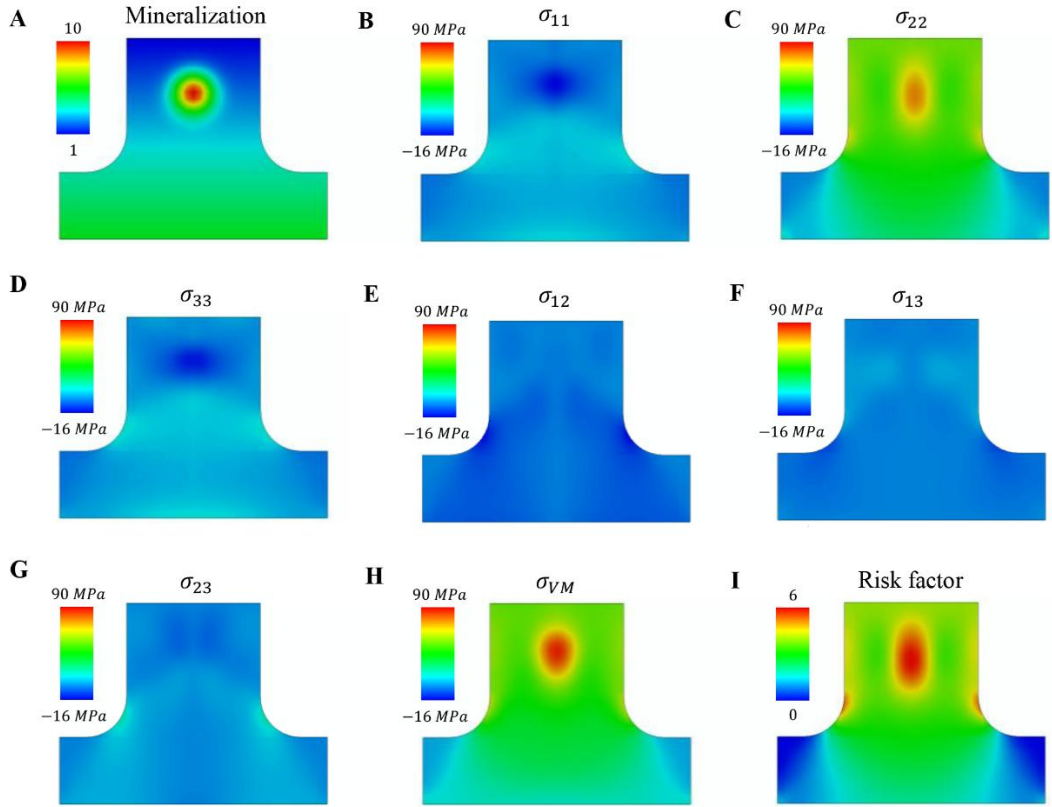

**Figure S2 | Stress and risk factor fields in the presence of localized mineralization.** A. Axisymmetric input mineralization scale field showing localized mineralization. B-I. Axisymmetric contours of $\sigma_{11}$, $\sigma_{22}$, $\sigma_{33}$, $\sigma_{12}$, $\sigma_{13}$, $\sigma_{23}$, von Mises stress, and the risk factor, respectively, all shown in the local coordinate system.

$\sigma_{VM}$ also exhibits a pronounced concentration within the mineralized region, with magnitudes substantially exceeding those of the individual stress components, reflecting the combined contributions and contrasts among the full stress state. More importantly, the proposed risk factor effectively captures the salient features of the stress fields in the presence of local mineralization. It not only reproduces the sharp concentration within the mineralized region but also reveals the enlarged and more pronounced elevations associated with pathological heterogeneity. In addition, the risk factor highlights secondary stress concentrations at the curved surfaces, consistent with the distributions observed in individual stress components. Together, these observations demonstrate that the risk factor provides a robust and physiologically relevant metric for assessing fibril-level failure susceptibility, even under pathological conditions characterized by localized mineralization.

Taken together, these results demonstrate that localized mineralization dramatically amplifies stress concentrations and substantially elevates the risk of failure in tendon–bone interface fibers. This finding highlights the critical mechanical consequences of mineralization heterogeneity in pathological states and underscores the importance of maintaining graded, rather than abrupt or highly localized, mineralization profiles to preserve the structural integrity and mechanical resilience of soft–hard tissue interfaces.

## 4. Training of the CNNFP

Two separate CNNFPs are trained to predict the risk factor field and the normal stress component $\sigma_{22}$, respectively. Training is performed using the datasets described in the Methods section, with 1200 samples used for training and 400 samples reserved for validation. The evolution of the training and validation losses over the course of the training process is shown in Fig. S3A for the risk factor predictor and Fig. S3B for the $\sigma_{22}$ predictor.

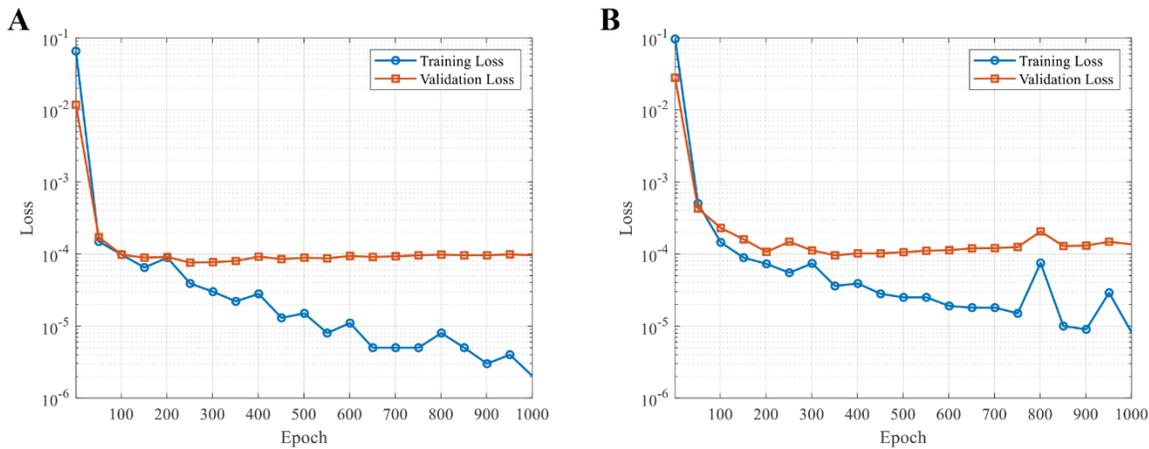

**Figure S3 | Training and validation losses plotted against the number of epochs.** A. CNNFP for the risk factor. B. CNNFP for $\sigma_{22}$.

For both CNNFPs, the training and validation losses decrease rapidly during the early stages of training and subsequently converge to stable values on the order of $10^{-4}$. The close agreement between training and validation loss curves indicates minimal overfitting and demonstrates strong generalization performance on unseen data. The smooth convergence behavior further suggests that the selected network architectures, loss function, and optimization strategy are well-suited for learning the nonlinear mapping between high-dimensional input fields and the corresponding mechanical response fields.

These results confirm that the trained CNNFPs provide accurate and reliable surrogates for finite element simulations, forming a robust foundation for subsequent field prediction, optimization, and inverse design tasks.

## 5. Randomly Generated Input Fields for Fig. 6A

Fig. S4 shows the randomly generated input fields of mineralization scale, angular dispersion, and the $x'$, $y'$, and $z'$ components of mean fiber orientation that lead to the risk factor field shown in Fig. 6A.

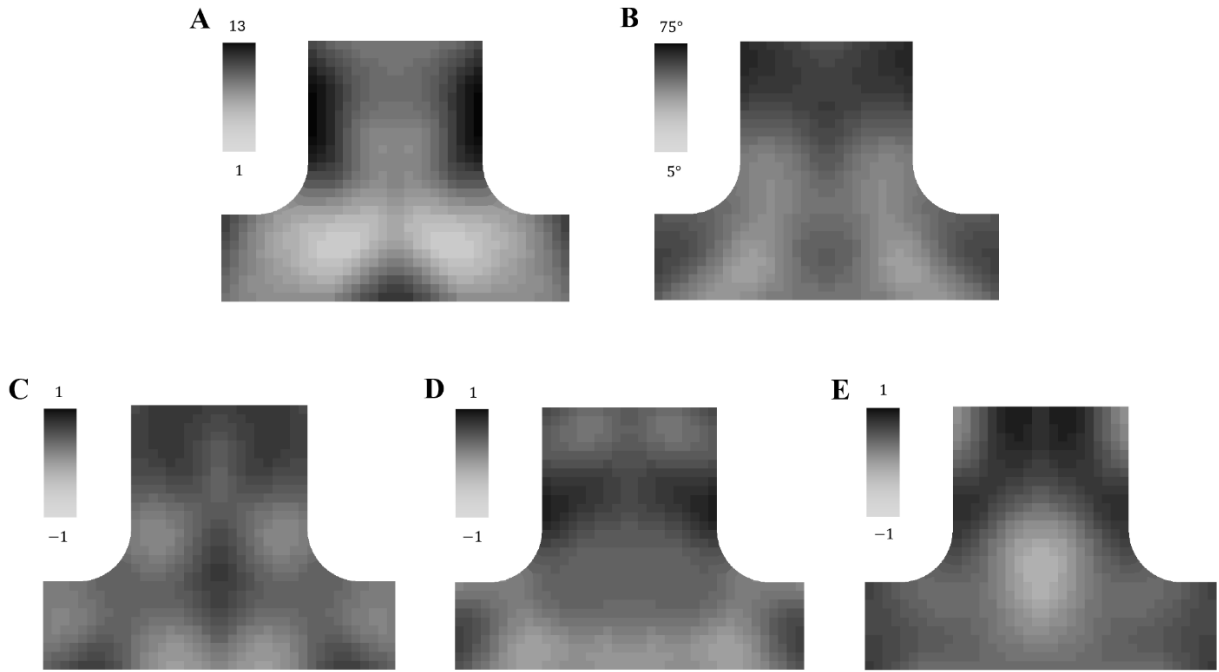

**Figure S4 | The randomly generated input fields related to the risk factor field shown in Fig. 6A. A-E.** Optimized input fields of mineralization scale, angular dispersion, and the $x'$, $y'$, and $z'$ components of mean fiber orientation, respectively.

## 6. Kernel-based Gradient Descent for Field Optimization

Before parameterization for optimization, the spatial input fields are first normalized such that their values are confined to the interval [0, 1]. The normalization is performed according to Eq. (S19).

$$\bar{f}^{(l)} = (f^{(l)} - F_{min}^{(l)})/(F_{max}^{(l)} - F_{min}^{(l)}) \quad (l = 1, 2, 3, 4, 5) \quad (S19)$$

where $F_{max}^{(l)}$ and $F_{min}^{(l)}$ denote the global upper and lower bounds of fields $f^{(l)}$, respectively. Their corresponding values are listed in Table S4.

**Table S4**. Global upper and lower bounds of the spatial input fields considered in this study.

| Field Name | $l$ | Upper Bounds $F_{max}^{(l)}$ | Lower Bounds $F_{min}^{(l)}$ |
|---|---|---|---|
| Mineralization scale | 1 | 13 | 1 |
| Angular dispersion | 2 | 75° | 5° |
| $x'$ component of mean fiber orientation | 3 | 1 | -1 |
| $y'$ component of mean fiber orientation | 4 | 1 | 0 |
| $z'$ component of mean fiber orientation | 5 | 1 | -1 |

Instead of optimizing every element of the normalized field $\bar{f}^{(l)}$ (i.e., individual image pixels), we optimize the kernel nodal values ($\alpha_i^{(l)}$ in Eq. (6)). This approach is computationally more efficient and inherently preserves the smoothness of the optimized field. In this study, 121 kernel nodes are employed, and their coordinates are provided in Table S5. In the kernel function $k((r, y), (r_i, y_i))$, the parameter $\sigma$, which defines the spatial extent of the kernel, is set to 18 for all fields throughout this study.

**Table S5**. Spatial coordinates of the kernel nodes used in this study.

| Kernel Node ID | | $r$ | | | | | | | | | |
|---|---|---|---|---|---|---|---|---|---|---|---|
| | | -75 | -55 | -35 | -15 | 5 | 25 | 45 | 65 | 85 | 105 | 125 |
| | 125 | 1 | 2 | 3 | 4 | 5 | 6 | 7 | 8 | 9 | 10 | 11 |
| $y$ | 105 | 12 | 13 | 14 | 15 | 16 | 17 | 18 | 19 | 20 | 21 | 22 |
| | 85 | 23 | 24 | 25 | 26 | 27 | 28 | 29 | 30 | 31 | 32 | 33 |

| | 65 | 34 | 35 | 36 | 37 | 38 | 39 | 40 | 41 | 42 | 43 | 44 |
|---|---|---|---|---|---|---|---|---|---|---|---|---|
| | 45 | 45 | 46 | 47 | 48 | 49 | 50 | 51 | 52 | 53 | 54 | 55 |
| | 25 | 56 | 57 | 58 | 59 | 60 | 61 | 62 | 63 | 64 | 65 | 66 |
| | 5 | 67 | 68 | 69 | 70 | 71 | 72 | 73 | 74 | 75 | 76 | 77 |
| | -15 | 78 | 79 | 80 | 81 | 82 | 83 | 84 | 85 | 86 | 87 | 88 |
| | -35 | 89 | 90 | 91 | 92 | 93 | 94 | 95 | 96 | 97 | 98 | 99 |
| | -55 | 100 | 101 | 102 | 103 | 104 | 105 | 106 | 107 | 108 | 109 | 110 |
| | -75 | 111 | 112 | 113 | 114 | 115 | 116 | 117 | 118 | 119 | 120 | 121 |

We slightly modify *Adam* [21], a stochastic gradient-based optimization method, to optimize the kernel nodal value vector $\theta$, where $\theta[121(l-1)+i] = \alpha_i^{(l)}$ ($l = 1,2,3,4,5$ & $i = 1,2,\ldots,121$). Specifically, $\theta$ is updated based on Eq. (S20).

$$\theta_t = \min\left[0.4, \max\left(-0.4, \theta_{t-1} - \gamma \frac{\frac{m_t}{1-\beta_1^t}}{\sqrt{\frac{v_t}{1-\beta_2^t}} + \epsilon}\right)\right] \tag{S20}$$

where $\theta_t$ is the kernel nodal value vector at time step $t$, $\gamma$ is the gradient descent step size, $m_t$ and $v_t$ are first and second moment estimates, respectively, $\beta_1$ and $\beta_2$ are exponential decay rates for the first and second moment estimates, respectively, and $\epsilon$ is a small constant for numerical stability. In this study, we set $\gamma = 0.05$, $\beta_1 = 0.9$, $\beta_2 = 0.999$, and $\epsilon = 10^{-7}$. The first and second moment estimates are defined in Eq. (S21) and (S22), respectively.

$$m_t = \beta_1 m_{t-1} + (1-\beta_1) g_t \tag{S21}$$

$$v_t = \beta_2 v_{t-1} + (1-\beta_2) g_t^2 \tag{S22}$$

where $g_t$ denotes the stochastic gradient at time step $t$, as defined in Eq. (23).

$$g_t = \nabla_\theta f_t(\theta_{t-1}) \tag{S23}$$

$f_t(\theta_{t-1})$ is the loss at time step $t$, evaluated using the kernel nodal value vector from the previous time step. The loss is defined as

$$f_t(\theta_{t-1}) = \max\left(\mathcal{H}(\bar{f}^{(1)}(r,y;\theta_{t-1}), \bar{f}^{(2)}(r,y;\theta_{t-1}),\right.$$
$$\left.\bar{f}^{(3)}(r,y;\theta_{t-1}), \bar{f}^{(4)}(r,y;\theta_{t-1}), \bar{f}^{(5)}(r,y;\theta_{t-1}))\right) \tag{S24}$$

where $\mathcal{H}$ denotes the mapping from five input fields to a single output field learned by the CNNFP.

The kernel nodal value vector $\theta$ is initialized to zero, after which the optimization proceeds until either the optimization goal is achieved or the maximum number of iterations (MNOI), set to 300, is reached.